\documentclass[12pt]{article}
\input epsf.tex
\begin{document}

\bigskip
\begin{titlepage}

\vspace{1cm}

\begin{center}
{\Large\bf A general initial condition of inflationary cosmology
on trans-Planckian physics\\}

\end{center}
\vspace{3mm}

\begin{center}

{\large Ke Ke

\vspace{5mm}

\textit {Institute of Theoretical Physics\\
Academia Sinica, P.O.Box 2735\\
Beijing 100080, P.R. China}}

\vspace{3mm}

{\tt kek@itp.ac.cn\\}

\end{center}

\vspace{5mm}

\begin{center}
{\large \bf Abstract}
\end{center}
We consider a more general initial condition satisfying the
minimal uncertainty relationship. We calculate the power spectrum
of a simple model in inflationary cosmology. The results depend on
perturbations generated below a fundamental scale, e.g. the Planck
scale.


\vfill
\begin{flushleft}
December 2003
\end{flushleft}
\end{titlepage}
\newpage

In recent years it has been realized that tiny quantum
fluctuations can be magnified by inflation. So we can study high
energy physics through the cosmological observation. The results
from WMAP will give us more precise cosmological parameters and
the information which we can't get at the accelerators. At the
very high energy scale, general relativity will break down and
quantum gravity effects will appear. So we can get some
information of trans-Planckian physic through cosmology.

The trans-Planckian problems in inflation was first raised
explicitly in\cite{Brandenberger:Proceeding}. In general, there
are three ways to tackle this problem. One is to modify the
dispersion relations\cite{ph/0005432}-\cite{th/0005209}. In this
model the normal linear dispersion relation is replaced by new
dispersion relations which differ from the linear one on length
scales smaller than the Planck length.

Another line of approach to the trans-Planckian problems is to
modify the equation of motion\cite{th/0203119}-\cite{ph/0311378}.
Inspired from string theory, space-time is noncommutative. So the
space-space noncommutativity and the space-time noncommutativity
was studied to get the modified power spectrum of cosmological
fluctuations.

In yet another approach to the trans-Planckian issue,
Danielsson\cite{th/0203198}suggested the modified initial
condition. In this model, the modes representing cosmological
fluctuations are generated mode by mode at the time when the
physical wavelength of the mode equals to the Planck length, or
more generally, when the energy scale equals to a new physics
scale. This kind of vacuum states are called $\alpha$ vacua. This
model was discussed in many
papers\cite{th/0205227}-\cite{ph/0307011}, the linear order
corrections in H/$\Lambda$ of power spectrum is got, where $H$ is
the Hubble constant.

In this paper, we assume that some fluctuations are already
generated below the Planck scale and it is a state satisfying the
minimum uncertainty relationship as the vacuum. This kind of
quantum fluctuations can also be magnified by inflation. So the
power spectrum of the cosmological fluctuations will be modified
by this choice of initial condition.

Let us consider a simple case in de-Sitter space-time, the metric
of space-time is
\begin{equation}
ds^{2}=-a\left( \eta\right)^{2}\left( d\eta^{2}-dx^{2}\right)
\end{equation}
where the $\eta$ is the conformal time and $ a\left( \eta\right)=
-\frac{1}{\eta H}$. Consider a scalar field in this background,
the action is
\begin{equation}
S=-\frac{1}{2}\int
d^{4}x\sqrt{-g}\partial_{\mu}\phi\partial^{\mu}\phi
\end{equation}
rescale the scalar field with $\mu=a\phi$, we get the action in
$\bf {k}$ space
\begin{eqnarray}
S=-\frac{1}{2}\int d\eta \int
d^{3}{\bf{k}}\biggl[\mu_{\bf{k}}^{\prime}\mu_{\bf{k}}^{\dagger
\prime}+\frac{a^{\prime
2}}{a^{2}}\mu_{\bf{k}}\mu_{\bf{k}}^{\dagger}-\frac{a^{\prime}}{a}(\mu_{\bf{k}}^{\prime}\mu_{\bf{k}}^{\dagger}+\mu_{\bf{k}}\mu_{\bf{k}}^{\dagger
\prime})-k^{2}\mu_{\bf{k}}\mu_{\bf{k}}^{\dagger}\biggr]
\end{eqnarray}
and the Hamitonian
\begin{eqnarray}
H=\frac{1}{2}\int
d^{3}{k}\biggl[\pi_{\bf{k}}^{\prime}\pi_{\bf{k}}^{\dagger
\prime}+\frac{a^{\prime}}{a}(\mu_{\bf{k}}\pi_{\bf{k}}^{\dagger}+\mu_{\bf{k}}^{\dagger}\pi_{\bf{k}})+k^{2}\mu_{\bf{k}}\mu_{\bf{k}}^{\dagger}\biggr]
\end{eqnarray}
where
\begin{equation}
\pi_{\bf{k}}=\mu_{\bf{k}}^{\prime}-\frac{a^{\prime}}{a}\mu_{\bf{k}}.
\end{equation}
using the time dependent oscillators, we can write
\begin{eqnarray}
\mu_{\bf{k}}\left(\eta\right)&=&\frac{1}{\sqrt{2k}}\left(
a_{\bf{k}}\left(\eta\right)+a_{-\bf{k}}^{\dagger}\left(\eta\right)\right)   \nonumber \\
\pi_{\bf{k}}\left(\eta\right)&=&-i\sqrt{\frac{k}{2}}\left(a_{\bf{k}}\left(\eta\right)-a_{-\bf{k}}^{\dagger
}\left(\eta\right)\right)
\end{eqnarray}
The Hamitonian reads
\begin{eqnarray}
H=\int
d^{3}{\bf{k}}\biggl[ka_{\bf{k}}^{\dagger}a_{\bf{k}}+\frac{i}{2}\frac{a^\prime}{a}
\biggl(a_{-{\bf{k}}}^{\dagger}a_{\bf{k}}^{\dagger
}-a_{-{\bf{k}}}a_{\bf{k}}\biggr)\biggr]
\end{eqnarray}
The creation and annihilation operators satisfy the Heinsenberg
equations:
\begin{eqnarray}
i\frac{d}{d\eta}a_{\bf{k}}(\eta)&=&[a_{\bf{k}},H]   \nonumber \\
i\frac{d}{d\eta}a_{\bf{k}}(\eta)^{\dagger}&=&[a_{\bf{k}}^{\dagger},H]
\end{eqnarray}
The time dependence of the oscillators can be written
\begin{eqnarray}
a_{\bf{k}}\left(\eta\right)&=&u_{k}\left(\eta\right)a_{\bf{k}}\left(\eta
_{0}\right)+v_{k}\left(\eta\right)a_{-\bf{k}}^{\dagger}\left(\eta
_{0}\right)   \nonumber \\
a_{-\bf{k}}^{\dagger}\left(\eta\right)&=&u_{k}^{\ast}\left(\eta
\right)a_{-\bf{k}}^{\dagger}\left(\eta_{0}\right)+v_{k}^{\ast
}\left(\eta\right)a_{\bf{k}}\left(\eta _{0}\right)
\end{eqnarray}
where $\eta_{0}$ is some fixed initial time, e.g. the time
associated with the Planck energy. Plugging this back into the
Heinsenberg equations, we can get the solutions
\begin{eqnarray}
u_{k}&=&\frac{1}{2}\left(A_{k}e^{-ik\eta }\left(2-\frac{i}{k\eta
}\right)
+B_{k}e^{ik\eta}\frac{i}{k\eta}\right)   \nonumber \\
v_{k}^{\ast}&=&\frac{1}{2}\left( B_{k}e^{ik\eta }\left( 2+\frac{i}{k\eta }%
\right) -A_{k}e^{-ik\eta }\frac{i}{k\eta }\right) .
\end{eqnarray}
where
\begin{eqnarray}
A_{k}&=&(1+\frac{i}{2k\eta_{0}})e^{ik\eta_{0}} \nonumber\\
B_{k}&=&\frac{i}{2k\eta_{0}}e^{-ik\eta_{0}}
\end{eqnarray}
Now, let us choose the initial condition. In\cite{th/0203198}, the
choice is that the initial state is vacuum when $\eta=\eta_{0}$.
This choice means that there are no fluctuations at all until the
Planck scale. However, the physics is complex at Planck scale in
string theory as we know. So the initial may be a more complex
state but vacuum. If we assume that some quantum fluctuations are
generated below the Planck scale and keep the minimum quantum
fluctuations before inflation:
\begin{equation}
\Delta\mu_{k}\Delta\pi_{k}=\frac{1}{2}
\end{equation}
where
\begin{eqnarray}
\Delta\mu_{k}^{2}&=&<\mu_{\bf{k}}^{\dagger}\mu_{\bf{k}}>-<\mu_{\bf{k}}^{\dagger}><\mu_{\bf{k}}> \nonumber\\
\Delta\pi_{k}^{2}&=&<\pi_{\bf{k}}^{\dagger}\pi_{\bf{k}}>-<\pi_{\bf{k}}^{\dagger}><\pi_{\bf{k}}>
\end{eqnarray}
Except for vacuum, the only choice is coherent state.
\begin{equation}
a_{\bf{k}}(\eta_{0})\left|\alpha_{\bf{k}},\eta_{0}\right\rangle
=\alpha_{\bf{k}}\left| \alpha_{\bf{k}},\eta _{0}\right\rangle
\end{equation}
where
\begin{equation}
\alpha_{\bf{k}}=\sqrt{N_{k}}e^{i\theta_{\bf{k}}}
\end{equation}
$N_{k}$ is the particle number representing the amplitude of the
fluctuations and $\theta_{\bf{k}}$ is the phase factor
representing the phase of the fluctuations. The Power spectrum is
\begin{eqnarray}
P_{\phi}&=& \frac{k^{3}}{2\pi^{2}a^{2}}\left<\mu_{\bf{k}}^{\dagger}\mu_{\bf{k}}\right>   \nonumber\\
&=&\frac{k^{2}}{4\pi^{2}a^{2}}\left<
\biggl(\left(u_{k}+v_{k}^{*}\right)a_{\bf{k}}+\left(u_{k}^{*}+v_{k}\right)a_{-\bf{k}}^{\dagger}\biggr)\biggl(\left(u_{k}^{*}+v_{k}\right)a_{\bf{k}}^{\dagger}+\left(u_{k}+v_{k}^{*}\right)a_{-\bf{k}}\biggr)
\right>
\end{eqnarray}
For $\eta \rightarrow 0$, $v_{k}^{\ast}(\eta)\rightarrow
u_{k}(\eta)$, the spectrum reads
\begin{eqnarray}
P_{\phi}=\frac{k^{2}}{\pi^{2}a^{2}}\left|u_{k}\right|^{2}\biggl[1+N_{k}\left(2+e^{i(2\theta(\eta)+\theta_{\bf{k}}+\theta_{-\bf{k}})}+e^{-i(2\theta(\eta)+\theta_{\bf{k}}+\theta_{-\bf{k}})}\right)\biggr]
\end{eqnarray}
where $\theta(\eta)$ is the phase of $u_{k}$,
$u_{k}=\left|u_{k}\right|e^{i\theta(\eta)}$. In bracket, the first
term represents the fluctuations generated after $\eta_{0}$, the
second term proportional to $N_{k}$ represents the earlier quantum
fluctuations magnified by inflation. It is proportional to
$<\mu_{\bf{k}}^{\dagger}><\mu_{\bf{k}}>$. If the phase factors of
the initial coherent state satisfy
$\theta_{\bf{k}}+\theta_{-\bf{k}}=-2\theta(\eta)+\pi$, it implies
$<\mu_{k}>=0$ at late times. We can see that the correction
proportional to $N_{k}$ totally vanishes. So we can't see the
fluctuations generated below Planck scale. But this condition
seems to be not reasonable. Why we should choose the initial phase
factor at $\eta=\eta_{0}$ satisfying $<\mu_{\bf{k}}>$ at late
times? Now let us consider the Hamitonian of a fixed $\bf{k}$
mode. At $\eta=\eta_{0}$, the energy is
\begin{equation}
E|_{\eta=\eta_{0}}=\left<H|_{\eta=\eta_{0}}\right>=kN_{k}\left[1-\frac{1}{k\eta_{0}}\rm{sin}(\theta_{\bf{k}}+\theta_{-\bf{k}})\right]
\end{equation}
The same as \cite{th/0203198}, where
$k\eta_{0}=-\frac{\Lambda}{H}$.
\begin{equation}
E|_{\eta=\eta_{0}}=kN_{k}\left[1+\frac{H}{\Lambda}\rm{sin}(\theta_{\bf{k}}+\theta_{-\bf{k}})\right]
\end{equation}
So the more reasonable initial condition is
$\theta_{\bf{k}}+\theta_{-\bf{k}}=\frac{3}{2}\pi$, which minimize
the energy at $\eta=\eta_{0}$. This condition also set
$<\mu_{\bf{k}}>=0$ at $\eta=\eta_{0}$. Plugging this initial
condition into (17), we can get
\begin{equation}
P_{\phi}=\frac{k^{2}}{\pi^{2}a^{2}}\left|u_{k}\right|^{2}\biggl[1+2N_{k}\left(1+\rm{sin}2\theta(\eta)\right)\biggr]
\end{equation}
Expanding the result in first order of $\frac{H}{\Lambda}$, we can
get
\begin{equation}
P_{\phi}=\biggl(\frac{H}{2\pi}\biggr)^{2}\biggl[1-\frac{H}{\Lambda}{\rm{sin}}(\frac{2\Lambda}{H})+2N_{k}(1+{\rm{sin}}\frac{2\Lambda}{H}-\frac{H}{\Lambda}{\rm{sin}}\frac{2\Lambda}{H}+\frac{H}{\Lambda}{\rm{cos}}\frac{2\Lambda}{H}-\frac{H}{\Lambda})\biggr]
\end{equation}

Some comments on this results are in order. First, the power
spectrum depends on the phase factor of the initial state. As we
know, most of the physical observable quantities are independent
of the phase factor in quantum mechanics. But we can see that the
phase factor will affect the power spectrum. Different choices of
phase factor result in different corrections.

Second, the first order correction in \cite{th/0203198}
proportions to ${\rm{sin}}\frac{2\Lambda}{H}$. Because
$\frac{\Lambda}{H}$ is a very large number, the different choices
of $\Lambda$ result in a big oscillations in correction terms. It
seems unphysical. In our results, we get a correction term
proportional to $N_{k}$, it is independent of the choice of
$\Lambda$. Interestingly, If we adjust the
$N_{k}=\frac{H}{2\Lambda}$, we can get the linear correction of
$\frac{H}{\Lambda}$.
\begin{equation}
P_{\phi}=\biggl(\frac{H}{2\pi}\biggr)^{2}\biggl(1+\frac{H}{\Lambda}\biggr)
\end{equation}
But we don't know how to determine $N_{k}$ which represent the
magnitude of the fluctuations below Planck scale. It may rely on
the details of high-energy physics.

Now, a short conclusion can be made. In this paper, we assume that
some quantum fluctuations are generated below the new physics
scale. Because we know little about the high-energy physics, we
assume that all of high-energy physics information is imprinted
effectively in the initial state. To maintain the minimal
uncertainty relationship, we set the coherent state as the initial
state. This kind of quantum fluctuations also can be magnified by
inflation. The result depend on the eigenvalue of the coherent
state. If the fluctuations generated below Planck scale are big
enough, i.e. $N_{k}$ is big enough, the effect of trans-Planckian
physics are possibly observed by more precise cosmological
observation. It is possibly a window through which we can see the
quantum gravity information.

\section*{Acknowledgments}

\bigskip

I am particularly grateful to Miao Li for his valuable discussions
and suggestions. I also thank Qing-Guo Huang and Peng Zhang for
their discussions.

\end{document}